\documentclass[11pt,a4paper]{article}
\usepackage[hyperref]{acl2018}
\usepackage{times}
\usepackage{latexsym}
\usepackage{graphicx}

\usepackage{url}

\aclfinalcopy % Uncomment this line for the final submission
\title{Sequence to Sequence Learning for  Query Expansion}

\author{Mohamed Salah Zaiem	\\
  Ecole Polytechnique \\
  {\tt   mohamed-salah.zaiem  } \\
  @polytechnique.edu \\\And
  Fatiha Sadat \\
  Universite du Quebec a Montreal \\
  {\tt sadat.fatiha@uqam.ca} \\}

\date{}

\begin{document}
\maketitle
\begin{abstract}
As fas as we are aware, using sequence to sequence algorithms for query expansion has not been explored yet in Information Retrieval literature nor in Question-Answering's. We tried to fill this gap in the literature with a custom Query Expansion system trained and tested on open datasets. One specificity of our engine compared to classic ones is that it does not need the documents to expand the introduced query. We test our expansions on three different tasks : Information Retrieval, Answer preselection and Text classification. Our method yielded a slight improvement in performance in the three tasks . 
\end{abstract}

\section{Introduction}
Recent works on Search-Oriented chatbots have been focusing on the reranking process after the preselection of a small number (usually around 10) of similar queries or hypothetical answers. Using basic similarity measures (Bag of words vectorization using BM25 or TF-IDF weighting for instance), a few hypotheses are selected for further inspection. 
\\
The preselection phase is rarely tackled in the literature. It is nevertheless an important task. Two reasons illustrate its importance. First, missing the correct answer or document in the preselection caps the accuracy at a certain coverage. Second, finding the relevant document in the fewest number of hypotheses makes the reranking process faster by reducing the number of answers to rerank. With the reranking processes getting more and more complex \cite{zhao:2016,qiu:2017} while the need is for faster answer generation, a fast and efficient preselection increases the accuracy and improves the users' experience.
\\
The importance of Query Expansion in Question Answering is highlighted in \cite{derczynski:2008}. To avoid the knock-on effect induced by a failure in any steps of the QA system, Query Expansion through Pseudo Relevance Feedback (PRF) is used to improve the IR component performance. PRF improves the coverage of the Answer Extraction component.

Throughout this work, we will explore a new technique using Sequence to Sequence neural architecture to expand the queries introduced by the users. 
\\
Here is a summary of our main contributions : 

\begin{itemize}

\item Starting from open datasets, we built a Query Expansion training set using sentence-embeddings-based Keyword Extraction.
\item We assess the ability of the Sequence to Sequence neural networks to capture expanding relations in the words embeddings' space. 

\end{itemize} 
Our work is organized as follows : Section 2 presents the state-of-the-art related to this research. Section 3 presents our approach. Section 4 shows our experiments and  and evaluations. Section 5 concludes this paper and presents some perspectives.
\section{Related Work }

Relevance feedback has been a popular choice for query expansion, starting with the Rocchio Algorithm \cite{salton:1971} in SMART Information Retrieval System. Using a set of relevant and non relevant documents, the original query vector is modified.  Pseudo Relevance Feedback relies on terms collected from the most pseudo-relevant documents of a first search. These terms are added to the query and a second search gives the final documents/answers. 
\\
But, we can also expand queries using external resources. Query expansion also relied on  anthologies and thesaurus-based methods, using large lexical databases like WordNet \cite{miller:95}. The query is expanded by adding terms using synonymy, hypernymy and other semantic relationships. \\

Recently, the introduction of word embeddings \cite{mikolov:2013} allowed new possibilities for Query Expansion \cite{roy:2016}. The distributed representations of the words in a query made it possible to produce expansions without extracting them from the documents. Using the centroid of the words introduced and cosine-similar tokens, Kuzi and al \shortcite{kuzi:2016} proposed a document-independent expansion method. This method has been therefore integrated to a custom PRF technique yielding particularly interesting results. As referenced by Mitra and Crasswell \shortcite{mitra:2017} , Diaz and al \shortcite{diaz:2016} showed that locally trained word embeddings along with topic specific language models lead to a significant improvement on Information Retrieval tasks.
\\

Lately, Nogueira and Cho \shortcite{nogueira:2017}  proposed a Reinforcement Learning based query expander. After selecting a set of candidate terms from the documents, a search engine looks for the relevant documents. Using relevance judgments, the system measures how much the suggested expansion improved the search accuracy and updates the parameters of the network. 
\\

Paraphrase generation is a very close field. Generating new utterances carrying the same meaning expands the initial query and highly increases  the robustness of a search-based chatbot \cite{mcclendon:2014,kozlowski:2003}.

Recently, Prakash and al. \shortcite{prakash:2016} proposed a model using stacked residual LSTM networks. While it reaches high BLEU and METEOR scores, we lack information about how different  the produced queries are from the introduced ones. And therefore, we cannot assess how much do they increase a conversational engine robustness. In addition, Buck and al. \shortcite{buck:2017} proposed a Reinforcement Learning based method for question reformulation. The reformulation network is updated according to the performance on Question Answering.
\\

\subsection{Sentence embeddings}
A vectorial representation of whole sentences and documents is useful for many tasks such as Text Classification.  Arora and al. \shortcite{arora:2017} proposed a simple baseline that performs well in several NLP tasks. It starts by averaging weighted pre-trained vectors of the words in the sentence. Then it deletes a common part to all sentences by removing a projection on the first singular vector of the sentences vectors' matrix. Another approach developed by Conneau and al \shortcite{Conneau:2017} used the SNLI database and a BiLSTM architecture to improve the sentences encoding using Natural Language Inference detection. The network updates its encoding of the sentences to predict the semantic relationship between two sentences. We will use a feature of that BiLSTM encoder to extract the keywords of a sentence in our expansion generation engine. \\
\subsection{Sequence to sequence architectures} 
Sequence to Sequence is a  neural architecture very popular in machine translation since it achieved state of the art results \cite{luong:2015}. Proposed by Sutskever and al \shortcite{Sutskever:2014} and Kalchbrenner \& Blunsom \shortcite{kalchbrenner:2013}, it consists in a two-component model using recurrent neural networks to link variable-length input sequences to variable-length output sequences. The introduced sequence gets encoded by the first component into a vectorial representation. Therefore, the decoder transforms that vector into the target sequence. For the example of automatic translation, the source sequence is a sentence in language A. The decoder outputs the target sentence in language B.
\\
The target sequence is the argmax of    $${p}(Y) = \prod\limits_{\scriptstyle k = 1\hfill}^T
{p(y_t |{y_1,...,y_{t-1}},c)} $$
And at each step the next token maximizes : 
$${p}(y_i|y_1,...,y_{i-1},x) = g(y_{i-1}, s_i, c) $$
where $s_i$ is the i-th hidden state of the encoder, c the final vector output by the encoder representing the entire input sentence and $y_i$ the i-th generated token. g is the function learned by the decoder. 
\\The encoding and decoding parts are usually composed of RNN cells. Bengio and al \shortcite{Bengio:1994} showed that Long Short Term Memory cells were very efficient to deal with vanishing gradient issues. LSTM cells \cite{schmidhbuber:1997} help the network capture long term dependencies and therefore exploit long sequences.
\\
Attention mechanisms are an extension to the encoder-decoder model. Proposed first by Bahdanau and al. \shortcite{Bahdanau:2014}, it helps at each step the decoder with a context vector pointing out where the relevant information about the next token is located.
With attention, the formula leading to the generation of the next token becomes : 
$${p}(y_i|y_1,...,y_{i-1},x) = g(y_{i-1}, s_i, c_i) $$
We remark that the main difference is that the vector c becomes dependent of the rank of the generated token i. The vector $c_i$ is a weighted sum of annotations $h_1,...,h_{T} $ used in the encoding process ($T$ being the length of the input sequence). Using hidden states, the vector $c_i$ points out how much a part of the source sequence should participate into the generation of the next token.

\section{Our approach}
\subsection{Building the training set}
\subsubsection{Databases} 
If a lot of paraphrasing databases exist and are made available for free, they usually consist in very short paraphrases (synonyms or slightly changed reformulations). These datasets are not very appropriate to enhance and expand question queries as they would tackle each token separately, while we are trying to find expansions based on whole sentences. The second case is very long paraphrases ( piece of news on the same special event ). They are too long to be trained for conversational queries.  This is why we had to find genuine sources of training material. 
\\
We used MultiNLI \cite{Williams:2017} and SNLI \cite{Bowman:2015}. NLI stands for Natural Language Inference and describes datasets presenting a list of sentence pairs with the semantic relationship linking the two sentences. The main semantic relationships are "Neutral" for approximate paraphrase, "Entailment" for semantic entailment (Example :  (A) The president was assassinated. entails (B) The president is dead. )and "Contradiction". For both corpuses, we naturally eliminate pairs classified as contradiction as they shall not provide relevant expansions.
\\
We also selected the duplicate pairs from the Quora question pairs dataset and trained our expansion model using the words that do not appear in the first formulation. \\
Finally, MSCOCO \cite{Lin:2014} dataset consists in  human annotated captions of over 120K images. Since they are describing the same image, (which usually focuses on only a few objects and generally one prominent object or action), we can assume the  words appearing in one description and not in the other are an eventual expansion for the first annotation. 
\begin{table}[h]
\begin{center}
\begin{tabular}{|l|rl|}
\hline \bf Dataset & \bf Size & \\ \hline
Stanford NLI & 570k  &\\
Multi NLI & 433K &\\
Quora Duplicates & 404k &\\
Images annotations MSCOCO  & 120K &\\

\hline
\end{tabular}
\end{center}
\caption{\label{font-table}Datasets }
\end{table}
\subsubsection{Keywords extraction}

We will use sentence embeddings to find out which words contribute most to the final vector. This computation is based on the hidden states of the encoder. The last layer in the Infersent model is a Maxpooling one. A max pooling layer performs down-sampling by dividing the input into rectangular pooling regions, and computing the maximum of each region. What we can see in the Figure 1 is the number of times the maxpooling layer chose the hidden state $h_t$ which is like a sentence representation centered around the t-th token. The words chosen the most times will be our selected keywords. These keywords will compose the expansions.  
\begin{figure}
\centering
\includegraphics[width=0.43\textwidth]{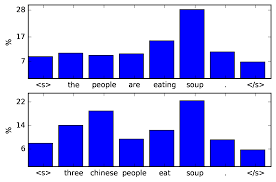}
\caption{\label{fig:conversations}Keyword extraction with the maxpooling layer (Credits : Conneau and al.)}
\end{figure}
Example : a picture of an old parade going through a town
gives these extracted keywords : 
['parade', 'old', 'picture', 'town']

\subsection{The training}
\subsubsection{Preprocessing}
The specificity of the Query Expansion task makes us able to take both A-B and B-A pairs for training. We extract the keywords from the target sequence, and then we remove the ones that appear in the source sequence. \\
To get targets with similar lengths, we remove the pairs with a target having less than 3 tokens. We also limit the number of target tokens to 6.\\
We finally get 520k pairs of sentence-expansion, this number may not be large enough for Sequence to Sequence learning, but we are limited by the number of available exploitable datasets. Here is an example from our training set :
\begin{quote}
Query :  who is the president of the U.S?
Expansion : american elected actual
\end{quote} 
 \subsubsection{Training Model}
We initiated the encoder and decoder weights with pre-trained word embeddings. We chose Glove \cite{Pennington:2014} 840B with 300 dimensioned vectors. \\
To choose the hyper parameters, we relied mainly on the best practices given by Britz and al. \shortcite{Britz:2017}. They conducted massive tests to give best practices advice for these architectures. We used a Bidirectional LSTM encoder with two layers of 500 hidden units. Graves and Schimdhuber \shortcite{graves:2005} have shown that Bidirectional encoding outperforms unidirectional one. The idea behind is to present the sentence in both forwards and backwards to two separate LSTM networks. It ensures that for every point in the input sequence, the network has complete information about the token preceding and following him.  The decoder is  a 2-layer LSTM with 500 hidden units.  \\
We used mini batches of 32 examples, and applied a 0.35 dropout probability in the LSTM stacks. We used Stochastic Gradient Descent as our optimizer and started with a learning rate of 0.001. The learning rate goes down with a decay of 0.5 after every epoch. There were 25 epochs of training for a total training time of 85 hours.
\\
The loss function is a Softmax cross entropy loss comparing the words computed by the decoder and the actual "True" targets. \\
\iffalse
It sadly does not take into account the words semantic similarity. A possible track for future work is a custom loss function taking into words' similarities using pre-trained word embeddings. We did not have enough time to explore this possibility. \\
\fi
We use a Bahdanau attention model for the expansion generation. We could think that attention is not needed in our context, as tokens should be produced based on the whole sentence. Yet, focusing on part of the sentences yields better results and was therefore our choice. \\
After generating the expanding sequence, we remove the words appearing in the initial query, and expand the source sentence with the remaining ones. 
The training accuracy reaches 25.85 percent. But it is not a relevant metric to test our query expansion model. Instead, using the same "search" component, we will evaluate the effect of the query expansion on the quality of the results/documents proposed. 
\\
\begin{figure}
\centering
\includegraphics[width=0.4\textwidth]{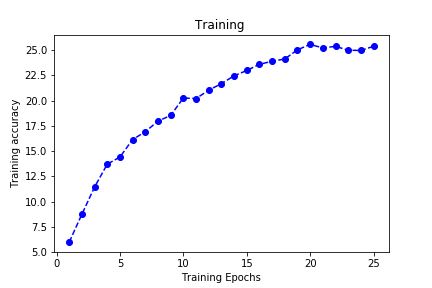}
\caption{\label{fig:conversations}Training evolution}
\end{figure}

\section{Evaluation}
We tested our query expansion model on three different tasks : Information Retrieval, Text Classification and Answer preselection. \\

\subsection{Information Retrieval}
For this task, we will use the TREC Robust 2004 Dataset \cite{voorhees:2001}. It consists in a set of 250 queries and 528,155 documents. For the search component, we will use Apache Lucene search.  \\
We start with the queries, and we expand them using our QE system and then we check the quality of the results provided by Lucene Search using StandardAnalyzer and two weighting schemes : BM25 and TF-IDF. \\ 
Here is an example :
\begin{itemize}
\item Query : bullying prevention programs /
Expansion : school program security
\end{itemize}
We will use Mean Average Precision as our metric to evaluate the results provided by our search. 
Mean Average Precision is the mean of the average precision scores for each query.
 $$MAP =\frac{\sum_{q=1}^{Q} AveP(q)}{Q} $$
 with Q the number of queries and  : 
 $${AveP} = \frac{\sum_{k=1}^n (P(k) \times {rel}(k))}{\mbox{number of relevant documents}} \! $$ 
 with n the number of documents, rel(k) an indicator function equaling 1 if the item at rank  k is a relevant document, zero otherwise and : 
 $$ \mbox{P}=\frac{|\{\mbox{relevant docs}\}\cap\{\mbox{retrieved docs}\}|}{|\{\mbox{retrieved docs}\}|} $$
The table shows the MAP results: 
\begin{table}[h]
\begin{center}
\begin{tabular}{|l|rl|}
\hline \bf Method & \bf MAP & \\ \hline
TF-IDF without QE & 0.2517  &\\
TF-IDF with QE & 0.2581&\\
BM25 without QE  & 0.2709&\\
BM25 with QE  & 0.2783 &\\

\hline
\end{tabular}
\end{center}
\caption{\label{font-table}Information Retrieval results }
\end{table}
\\
We run a Student t-test to check the significance of the difference. For the TF-IDF vectorization, the p-value reaches 0.434 while it equals 0.42 with BM-25. The difference is obviously not significant at 20\%  .
\subsection{Text Classification}

Text Classification is a very popular task in Natural Language Processing. To see how useful our QE engine can be for this task, we will use The Guardian API. We download a set of articles classified by topic from The Guardian newspaper. We will only use headlines for article classification. We divide the set, we take 40000 articles for training and we keep 1000 for testing. \\
For learning we use the SVM algorithm applied on vectors obtained with a basic TF-IDF vectorization. To make it easier, we only keep seven "large" labels : Culture, Sport, World, Politics, Business, Science and Media.\\
Our QE engine intervenes in the testing phase, we compare the results obtained with or without expanding the headlines of the testing set. We then compare the accuracies of the prediction : 

\begin{table}[h]
\begin{center}
\begin{tabular}{|l|rl|}
\hline \bf Method & \bf Accuracy & \\ \hline
Without QE & 0.7167  &\\
With QE & 0.7180 &\\

\hline
\end{tabular}
\end{center}
\caption{\label{font-table}Text Classification results }
\end{table}

We run the Student's T test on this task also. With a p-value of 0.47427, the difference is not significant at 20\% .

\subsection{Answer Preselection} 
This is the task that motivated us first to explore Query Expansion techniques.  \\
We will use the WikiQA Dataset \cite{yang:2015}. The WikiQA corpus is an open set of question and sentence pairs, collected and annotated for research on open-domain question answering. \\
For each question, we start a search on the set of answers with a similarity computation. We select the ten most similar answers, then we count the proportion of relevant answers in the ten hypotheses, taking into account that some questions have less than 10 possible answers. This will be our accuracy measure. Coverage is defined as  the proportion of queries that had at least one appropriate answer among the ten hypotheses. \\
We compare the accuracy measures using or not the Query Expansion engine. Here is an example of queries in the Dataset : 
\begin{itemize}
\item Query : How to lose weight ? /
Expansion : aerobic sport diet 
\end{itemize}
\begin{table}[h]
\begin{center}
\begin{tabular}{|l|r|l|}
\hline \bf Method & \bf  Accuracy & \bf  Coverage \\ \hline
TF-IDF without QE & 0.2871  & 0.7840\\
TF-IDF with QE & 0.2889 & 0.7901\\

\hline
\end{tabular}
\end{center}
\caption{\label{font-table}Answer Preselection results }
\end{table}
Due to the number of queries (2117), the t-test shows better results. However, with 0.44855 and 0.31393 p-values respectively for accuracy and coverage, the difference remains not significant at 20\%.
\section{Qualitative analysis and future tracks }
Our Query Expansion engine does improve the results in the different tasks we tested it in, but the progress is far from being impressive, and is logically not statistically significant. Here is a list of the issues limiting the reliability of our QE system and a few future tracks for improvement : 
\begin{itemize}
\item The QE system fails to capture the semantic mechanisms behind Query Expansion, and therefore could not expand queries of unseen topics. This may not be that surprising as the task seems very complicated and nothing proofs that the actual embedding space ensures and holds this type of semantic relationships. The expansions are therefore learned through the examples, and the models fails to enrich queries on topics it did not witness before (45\% of the queries are not expanded since no new word is added). The testing datasets are mainly composed of such queries. When we have a look to the attention matrices, we find out that the network does not rely on the question formulation in generating the new tokens. It almost only looks for the "keywords".
\item This makes us think that although it may not be efficient for open topics, training this model on local entailments would yield great expansion results. 
\item The nature of the testing queries, mainly the fact that they contained a lot of named entities, has been a huge handicap for our query expansions. Confronted to unknowns, the network replicates the words of the source sentence and the introduced query does not get any useful expanding word. We tried to remove the named entities in the sentence before expanding it. But as the sentence loses a lot of the information it carried first, the expansions were not very relevant.
\item We will explore the possibility of including the search in the training process. The progress on search would be a loss function updating the weights of the encoder-decoder network. After the first training, a second one ,based on the reward for the search, would refine the parameters of our network and make it more search-oriented.  
\end{itemize}
\section{Conclusion}
Throughout this work, we introduced a sequence to sequence framework for automatic query expansion. We implemented a genuine keyword extraction method to create the training dataset. When tested on three different tasks, our Query Expansion engines leads to a slight improvement. Although our model seems still unable to propose relevant expansions on unseen topics, it performs well on known ones. For the future, we are thinking about updating the Sequence to sequence network according to the search results. We would add on top of our encoder-decoder network a deep reinforcement learning component. This component would rely on the impact of our expansion on search quality.

% include your own bib file like this:
%\bibliographystyle{acl}
%\bibliography{acl2018}
\bibliography{mybib}

\begin{thebibliography}{}
\expandafter\ifx\csname natexlab\endcsname\relax\def\natexlab#1{#1}\fi

\bibitem[{Arora et~al.(2017)Arora, Liang, and Ma}]{arora:2017}
Sanjeev Arora, Yingyu Liang, and Tengyu Ma. 2017.
\newblock A simple but tough-to-beat baseline for sentence embeddings.
\newblock ICLR 2017.

\bibitem[{Bahdanau et~al.(2014)Bahdanau, Cho, and Bengio}]{Bahdanau:2014}
Dzmitry Bahdanau, Kyunghyun Cho, and Yoshua Bengio. 2014.
\newblock Neural machine translation by jointly learning to align and
  translate.
\newblock ICLR 2015.

\bibitem[{Bengio et~al.(1994)Bengio, Simard, and Frasconi}]{Bengio:1994}
Yoshua Bengio, Patrice Simard, and Paolo Frasconi. 1994.
\newblock \href{https://doi.org/10.1109/72.279181}{Learning long-term
  dependencies with gradient descent is difficult}.
\newblock {\em {IEEE} Transactions on Neural Networks\/} 5(2):157--166.
\newblock
  \href{https://doi.org/10.1109/72.279181}{https://doi.org/10.1109/72.279181}.

\bibitem[{Bowman et~al.(2015)Bowman, Angeli, Potts, and Manning}]{Bowman:2015}
Samuel~R. Bowman, Gabor Angeli, Christopher Potts, and Christopher~D. Manning.
  2015.
\newblock \href{https://doi.org/10.18653/v1/D15-1075}{A large annotated corpus
  for learning natural language inference}.
\newblock In {\em Proceedings of the 2015 Conference on Empirical Methods in
  Natural Language Processing\/}. Association for Computational Linguistics,
  pages 632--642.
\newblock
  \href{https://doi.org/10.18653/v1/D15-1075}{https://doi.org/10.18653/v1/D15-1075}.

\bibitem[{Britz et~al.(2017)Britz, Goldie, Luong, and Le}]{Britz:2017}
Denny Britz, Anna Goldie, Minh-Thang Luong, and Quoc Le. 2017.
\newblock \href{http://aclweb.org/anthology/D17-1151}{Massive exploration of
  neural machine translation architectures}.
\newblock In {\em Proceedings of the 2017 Conference on Empirical Methods in
  Natural Language Processing\/}. Association for Computational Linguistics,
  pages 1442--1451.
\newblock
  \href{http://aclweb.org/anthology/D17-1151}{http://aclweb.org/anthology/D17-1151}.

\bibitem[{Buck et~al.(2017)Buck, Bulian, Ciaramita, Gajewski, Gesmundo,
  Houlsby, and Wang}]{buck:2017}
Christian Buck, Jannis Bulian, Massimiliano Ciaramita, Wojciech Gajewski,
  Andrea Gesmundo, Neil Houlsby, and Wei Wang. 2017.
\newblock Ask the right questions: Active question reformulation with
  reinforcement learning.
\newblock Sixth International Conference on Learning Representations (ICLR),
  2018.

\bibitem[{Conneau et~al.(2017)Conneau, Kiela, Schwenk, Barrault, and
  Bordes}]{Conneau:2017}
Alexis Conneau, Douwe Kiela, Holger Schwenk, Lo{\"i}c Barrault, and Antoine
  Bordes. 2017.
\newblock \href{http://aclweb.org/anthology/D17-1070}{Supervised learning of
  universal sentence representations from natural language inference data}.
\newblock In {\em Proceedings of the 2017 Conference on Empirical Methods in
  Natural Language Processing\/}. Association for Computational Linguistics,
  pages 670--680.
\newblock
  \href{http://aclweb.org/anthology/D17-1070}{http://aclweb.org/anthology/D17-1070}.

\bibitem[{Derczynski et~al.(2008)Derczynski, Wang, Gaizauskas, and
  Greenwood}]{derczynski:2008}
Leon Derczynski, Jun Wang, Robert Gaizauskas, and Mark~A. Greenwood. 2008.
\newblock \href{http://aclweb.org/anthology/W08-1805}{A data driven approach to
  query expansion in question answering}.
\newblock In {\em Coling 2008: Proceedings of the 2nd workshop on Information
  Retrieval for Question Answering\/}. Coling 2008 Organizing Committee, pages
  34--41.
\newblock
  \href{http://aclweb.org/anthology/W08-1805}{http://aclweb.org/anthology/W08-1805}.

\bibitem[{Diaz et~al.(2016)Diaz, Mitra, and Craswell}]{diaz:2016}
Fernando Diaz, Bhaskar Mitra, and Nick Craswell. 2016.
\newblock \href{https://doi.org/10.18653/v1/P16-1035}{Query expansion with
  locally-trained word embeddings}.
\newblock In {\em Proceedings of the 54th Annual Meeting of the Association for
  Computational Linguistics (Volume 1: Long Papers)\/}. Association for
  Computational Linguistics, pages 367--377.
\newblock
  \href{https://doi.org/10.18653/v1/P16-1035}{https://doi.org/10.18653/v1/P16-1035}.

\bibitem[{Graves and Schmidhuber(2005)}]{graves:2005}
Alex Graves and Jürgen Schmidhuber. 2005.
\newblock \href{https://doi.org/10.1109/ijcnn.2005.1556215}{Framewise phoneme
  classification with bidirectional {LSTM} networks}.
\newblock In {\em Proceedings. 2005 {IEEE} International Joint Conference on
  Neural Networks, 2005.\/}. {IEEE}.
\newblock
  \href{https://doi.org/10.1109/ijcnn.2005.1556215}{https://doi.org/10.1109/ijcnn.2005.1556215}.

\bibitem[{Hochreiter and Schmidhuber(1997)}]{schmidhbuber:1997}
Sven Hochreiter and Jürgen Schmidhuber. 1997.
\newblock Long short-term memory.
\newblock Neural computation, volume~9, pages 1735--80.

\bibitem[{Kalchbrenner and Blunsom(2013)}]{kalchbrenner:2013}
Nal Kalchbrenner and Phil Blunsom. 2013.
\newblock \href{http://www.aclweb.org/anthology/D13-1176}{Recurrent continuous
  translation models}.
\newblock In {\em Proceedings of the 2013 Conference on Empirical Methods in
  Natural Language Processing\/}. Association for Computational Linguistics,
  pages 1700--1709.
\newblock
  \href{http://www.aclweb.org/anthology/D13-1176}{http://www.aclweb.org/anthology/D13-1176}.

\bibitem[{Kozlowski et~al.(2003)Kozlowski, McCoy, and
  Vijay-Shanker}]{kozlowski:2003}
Raymond Kozlowski, Kathleen~F. McCoy, and K.~Vijay-Shanker. 2003.
\newblock \href{http://www.aclweb.org/anthology/W03-1601}{Generation of
  single-sentence paraphrases from predicate/argument structure using
  lexico-grammatical resources}.
\newblock In {\em Proceedings of the Second International Workshop on
  Paraphrasing\/}.
\newblock
  \href{http://www.aclweb.org/anthology/W03-1601}{http://www.aclweb.org/anthology/W03-1601}.

\bibitem[{Kuzi et~al.(2016)Kuzi, Shtok, and Kurland}]{kuzi:2016}
Saar Kuzi, Anna Shtok, and Oren Kurland. 2016.
\newblock \href{https://doi.org/10.1145/2983323.2983876}{Query expansion using
  word embeddings}.
\newblock In {\em Proceedings of the 25th ACM International on Conference on
  Information and Knowledge Management\/}. ACM, New York, NY, USA, CIKM '16,
  pages 1929--1932.
\newblock
  \href{https://doi.org/10.1145/2983323.2983876}{https://doi.org/10.1145/2983323.2983876}.

\bibitem[{Lin et~al.(2014)Lin, Maire, Belongie, Bourdev, Girshick, Hays,
  Perona, Ramanan, Doll{\'{a}}r, and Zitnick}]{Lin:2014}
Tsung{-}Yi Lin, Michael Maire, Serge~J. Belongie, Lubomir~D. Bourdev, Ross~B.
  Girshick, James Hays, Pietro Perona, Deva Ramanan, Piotr Doll{\'{a}}r, and
  C.~Lawrence Zitnick. 2014.
\newblock \href{http://arxiv.org/abs/1405.0312}{Microsoft {COCO:} common
  objects in context}.
\newblock {\em CoRR\/} abs/1405.0312.
\newblock
  \href{http://arxiv.org/abs/1405.0312}{http://arxiv.org/abs/1405.0312}.

\bibitem[{Luong et~al.(2015)Luong, Pham, and Manning}]{luong:2015}
Thang Luong, Hieu Pham, and Christopher~D. Manning. 2015.
\newblock \href{https://doi.org/10.18653/v1/D15-1166}{Effective approaches to
  attention-based neural machine translation}.
\newblock In {\em Proceedings of the 2015 Conference on Empirical Methods in
  Natural Language Processing\/}. Association for Computational Linguistics,
  pages 1412--1421.
\newblock
  \href{https://doi.org/10.18653/v1/D15-1166}{https://doi.org/10.18653/v1/D15-1166}.

\bibitem[{McClendon et~al.(2014)McClendon, Mack, and Hodges}]{mcclendon:2014}
Jerome McClendon, Naja Mack, and Larry~F. Hodges. 2014.
\newblock The use of paraphrase identification in the retrieval of appropriate
  responses for script based conversational agents.
\newblock In {\em FLAIRS Conference\/}.

\bibitem[{Mikolov et~al.(2013)Mikolov, Sutskever, Chen, Corrado, and
  Dean}]{mikolov:2013}
Tomas Mikolov, Ilya Sutskever, Kai Chen, Greg~S Corrado, and Jeff Dean. 2013.
\newblock
  \href{http://papers.nips.cc/paper/5021-distributed-representations-of-words-and-phrases-and-their-compositionality.pdf}{Distributed
  representations of words and phrases and their compositionality}.
\newblock
  \href{http://papers.nips.cc/paper/5021-distributed-representations-of-words-and-phrases-and-their-compositionality.pdf}{http://papers.nips.cc/paper/5021-distributed-representations-of-words-and-phrases-and-their-compositionality.pdf}.

\bibitem[{Miller(1995)}]{miller:95}
George~A. Miller. 1995.
\newblock Wordnet: A lexical database for english.
\newblock Communications of the ACM, volume~38, pages 39--41.

\bibitem[{Mitra and Craswell(2017)}]{mitra:2017}
Bhaskar Mitra and Nick Craswell. 2017.
\newblock \href{http://arxiv.org/abs/1705.01509}{Neural models for information
  retrieval}.
\newblock {\em CoRR\/} abs/1705.01509.
\newblock
  \href{http://arxiv.org/abs/1705.01509}{http://arxiv.org/abs/1705.01509}.

\bibitem[{Nogueira and Cho(2017)}]{nogueira:2017}
Rodrigo Nogueira and Kyunghyun Cho. 2017.
\newblock \href{http://aclweb.org/anthology/D17-1061}{Task-oriented query
  reformulation with reinforcement learning}.
\newblock In {\em Proceedings of the 2017 Conference on Empirical Methods in
  Natural Language Processing\/}. Association for Computational Linguistics,
  pages 574--583.
\newblock
  \href{http://aclweb.org/anthology/D17-1061}{http://aclweb.org/anthology/D17-1061}.

\bibitem[{Pennington et~al.(2014)Pennington, Socher, and
  Manning}]{Pennington:2014}
Jeffrey Pennington, Richard Socher, and Christopher Manning. 2014.
\newblock \href{https://doi.org/10.3115/v1/D14-1162}{Glove: Global vectors for
  word representation}.
\newblock In {\em Proceedings of the 2014 Conference on Empirical Methods in
  Natural Language Processing (EMNLP)\/}. Association for Computational
  Linguistics, pages 1532--1543.
\newblock
  \href{https://doi.org/10.3115/v1/D14-1162}{https://doi.org/10.3115/v1/D14-1162}.

\bibitem[{Prakash et~al.(2016)Prakash, Hasan, Lee, Datla, Qadir, Liu, and
  Farri}]{prakash:2016}
Aaditya Prakash, Sadid~A. Hasan, Kathy Lee, Vivek Datla, Ashequl Qadir, Joey
  Liu, and Oladimeji Farri. 2016.
\newblock Neural paraphrase generation with stacked residual lstm networks.
\newblock COLING 2016.

\bibitem[{Qiu et~al.(2017)Qiu, Li, Wang, Gao, Chen, Zhao, Chen, Huang, and
  Chu}]{qiu:2017}
Minghui Qiu, Feng-Lin Li, Siyu Wang, Xing Gao, Yan Chen, Weipeng Zhao, Haiqing
  Chen, Jun Huang, and Wei Chu. 2017.
\newblock \href{https://doi.org/10.18653/v1/P17-2079}{Alime chat: A sequence to
  sequence and rerank based chatbot engine}.
\newblock In {\em Proceedings of the 55th Annual Meeting of the Association for
  Computational Linguistics (Volume 2: Short Papers)\/}. Association for
  Computational Linguistics, pages 498--503.
\newblock
  \href{https://doi.org/10.18653/v1/P17-2079}{https://doi.org/10.18653/v1/P17-2079}.

\bibitem[{Roy et~al.(2016)Roy, Paul, Mitra, and Garain}]{roy:2016}
Dwaipayan Roy, Debjyoti Paul, Mandar Mitra, and Utpal Garain. 2016.
\newblock Using word embeddings for automatic query expansion.
\newblock Neu-IR '16 SIGIR Workshop on Neural Information Retrieval.

\bibitem[{Salton(1971)}]{salton:1971}
G.~Salton. 1971.
\newblock {\em The SMART Retrieval System---Experiments in Automatic Document
  Processing\/}.
\newblock Prentice-Hall, Inc., Upper Saddle River, NJ, USA.

\bibitem[{Sutskever et~al.(2014)Sutskever, Vinyals, and Le}]{Sutskever:2014}
Ilya Sutskever, Oriol Vinyals, and Quoc~V. Le. 2014.
\newblock \href{http://dl.acm.org/citation.cfm?id=2969033.2969173}{Sequence to
  sequence learning with neural networks}.
\newblock In {\em Proceedings of the 27th International Conference on Neural
  Information Processing Systems - Volume 2\/}. MIT Press, Cambridge, MA, USA,
  NIPS'14, pages 3104--3112.
\newblock
  \href{http://dl.acm.org/citation.cfm?id=2969033.2969173}{http://dl.acm.org/citation.cfm?id=2969033.2969173}.

\bibitem[{Vorrhees(2001)}]{voorhees:2001}
Ellen~M. Vorrhees. 2001.
\newblock \href{https://doi.org/10.1017/s1351324901002789}{The {TREC} question
  answering track}.
\newblock Cambridge University Press ({CUP}), volume~7.
\newblock
  \href{https://doi.org/10.1017/s1351324901002789}{https://doi.org/10.1017/s1351324901002789}.

\bibitem[{Williams et~al.(2018)Williams, Nangia, and Bowman}]{Williams:2017}
Adina Williams, Nikita Nangia, and Samuel Bowman. 2018.
\newblock \href{http://aclweb.org/anthology/N18-1101}{A broad-coverage
  challenge corpus for sentence understanding through inference}.
\newblock In {\em Proceedings of the 2018 Conference of the North American
  Chapter of the Association for Computational Linguistics: Human Language
  Technologies, Volume 1 (Long Papers)\/}. Association for Computational
  Linguistics, pages 1112--1122.
\newblock
  \href{http://aclweb.org/anthology/N18-1101}{http://aclweb.org/anthology/N18-1101}.

\bibitem[{Yan et~al.(2016)Yan, Duan, Bao, Chen, Zhou, Li, and Zhou}]{zhao:2016}
Zhao Yan, Nan Duan, Junwei Bao, Peng Chen, Ming Zhou, Zhoujun Li, and Jianshe
  Zhou. 2016.
\newblock \href{https://doi.org/10.18653/v1/P16-1049}{Docchat: An information
  retrieval approach for chatbot engines using unstructured documents}.
\newblock In {\em Proceedings of the 54th Annual Meeting of the Association for
  Computational Linguistics (Volume 1: Long Papers)\/}. Association for
  Computational Linguistics, pages 516--525.
\newblock
  \href{https://doi.org/10.18653/v1/P16-1049}{https://doi.org/10.18653/v1/P16-1049}.

\bibitem[{Yang et~al.(2015)Yang, Yih, and Meek}]{yang:2015}
Yi~Yang, Wen-tau Yih, and Christopher Meek. 2015.
\newblock \href{https://doi.org/10.18653/v1/D15-1237}{Wikiqa: A challenge
  dataset for open-domain question answering}.
\newblock In {\em Proceedings of the 2015 Conference on Empirical Methods in
  Natural Language Processing\/}. Association for Computational Linguistics,
  pages 2013--2018.
\newblock
  \href{https://doi.org/10.18653/v1/D15-1237}{https://doi.org/10.18653/v1/D15-1237}.

\end{thebibliography}
\bibliographystyle{acl_natbib}

\end{document}